\documentclass{epl}
\title{Chiral Correction to the Spin Fluctuation Feedback in \\
two-dimensional $p$-wave Superconductors }
\author{J. Goryo\inst{1} \and M. Sigrist\inst{2,3}}
\institute{
  \inst{1} Institute for Solid State Physics, University of Tokyo,  
Kashiwanoha 5-1-5, Kashiwa, Chiba, 277-8581 Japan \\
  \inst{2} Yukawa Institute for Theoretical Physics, Kyoto University, Kyoto
606-8502, Japan \\
  \inst{3} Institute for Theoretical Physics, ETH-H\"onggerberg, 
8093 Z\"urich, Switzerland
}


\begin{document}

\maketitle

\begin{abstract}
We consider the stability of the superconducting phase for
spin-triplet $p$-wave pairing in a quasi-two-dimensional system. 
We show that in the absence of spin-orbit coupling there is a
chiral contribution to spin fluctuation feedback which is related
to spin quantum Hall effect in a chiral superconducting phase. 
We show that this mechanism supports the stability of a chiral
$p$-wave state. 
\end{abstract}

Cooper pairing in the spin-triplet (odd-parity) channel has been for
a long time the privilege of the superfluid $^3$He only \cite{HELIUM}. In the
mid-eighties superconductors have been discovered 
which were considered good candidates for spin-triplet pairing 
appeared, the so-called heavy Fermion superconductors, e.g. UPt$_3$ and
UBe$_{13}$ \cite{HF} More recently, several new compounds such as UGe$_2$ and
Sr$_2$RuO$_4$ have been added to the list of potential spin-triplet
superconductors. In particular, in the case the ruthenate 
we have to date overwhelming experimental evidence of spin-triplet
pairing in a state 
which resembles the A-phase of $^3$He from point of view of broken
symmetries \cite{Maeno}.

Spin-triplet pairing provides even in case of reduced rotation symmetry
(crystal field) still a large number of degrees of freedom, which 
leads in many cases to long lists of potential superconducting
phases \cite{REV}. An interesting situation occurs in a two-dimensional system,
where the weak-coupling approach leads to several spin-triplet pairing
states that possess the same condensation energy in the presence of
complete spin-rotation symmetry. Systems of this kind
are Sr$_2$RuO$_4$ which has a quasi-two-dimensional (2D) electron band
structure\cite{Maeno,Rice-Sigrist}  and thin films of $^3$He (whereby
we avoid the discussion of the Berezinsky-Kosterlitz-Thouless
transition which would occur in a film, see \cite{KORSH}). In both 
cases the state realized 
has the structure of a ``chiral $p$-wave'' state, with an orbital
angular momentum pointing out of the plane, $ p_x \pm i p_y $. This
chirality has been discussed in connection with a variety of possible 
effects, such as 
the zero-field Hall effect \cite{VOLOVIK,zero-field-Hall-effect},
modified vortex 
core states \cite{chiral-vor} or the spin quantum Hall effect (SQHE)
\cite{YAKO,Fisher,Read-Green}. In this letter we will show that 
chirality may also be an essential part to stabilization of the chiral
$p$-wave state through a spin-induced feedback effect.

We study here a 2D system with complete cylindrical rotation symmetry where
the gap function of the $p$-wave superconducting state can be
expressed as a $2 \times 2 $ -matrix $\Delta_{\alpha\beta}({\bf k})= 
i (\sigma_{\mu} \sigma_2)_{\alpha\beta} d_{\mu j} \hat{k}_j$ (
$d_{\mu j}$ is a complex order parameter ($\mu=1,2,3$, $j=x,y$) 
, $\sigma^{\mu}$ are the Pauli spin matrices (summation runs over
repeated indices) and $\hat{\bf k}={\bf k}/|{\bf k}| $). 
In the weak-coupling approach the condensation energy of a pairing
state depends only on the magnitude and shape of its quasi-particle gap, and
states with the same gap are degenerate. It is easy to see that the
most stable states in our case are the ones with 
an isotropic gap, $ | \Delta |^2 = d_{\mu i}^* d_{\mu j} \hat{k}_i
\hat{k}_j $. There are several different states
with the same isotropic gap which we may classify into two groups,

\begin{eqnarray}
|d_{\mu x}| = | d_{\mu y}| & \mbox{and} & Re(d^*_{\mu x}
 d_{\mu y}) = 0 \quad  \mbox{A-phase} \\
|d_{\mu x}| \neq | d_{\mu y}| & \mbox{and} & Im(d^*_{\mu x}
 d_{\mu y}) = 0 \quad  \mbox{B-phase}
\label{phases}
\end{eqnarray}
which are analogues of the A- (chiral) and B-phase of $^3$He,
respectively \cite{HELIUM,KORSH}. 
Obviously, both types of states possess an isotropic
gap. While spin-orbit coupling would lift the weak-coupling degeneracy
\cite{NG}, we ignore here this aspect and concentrate on the role of
the feedback effect. The feedback mechanism is based on the concept
that the modification of the pairing interaction caused by the
appearance of the superconducting condensate strengthens or
weakens a particular pairing channel. The spin fluctuation feedback
effect has been proposed as the mechanism for the stability of
the A-phase in superfluid $^3$He under pressure \cite{HELIUM,ABS-Kuroda}.
Recently, we have shown that for charged particles an additional,
though rather small, feedback effect exists which based on the
orbital chirality stabilizes the A-phase \cite{Goryo-Sigrist}. 
The problem of the weak-coupling 
degeneracy of the spin-triplet state is the motivation to revisit also
the spin fluctuation based feedback mechanism. 

We consider an electron system with a quasi-2D parabolic band $
\epsilon( {\bf k}) = \hbar^2 (k_x^2 + k_y^2 - k_{\rm F}^2)/2m$, 
where $k_{\rm F}$ is the Fermi wave number.  We
introduce the following spin-dependent two-particle interaction 
\begin{eqnarray} 
H_{int}&=& \frac{1}{2 \Omega} \sum_{{\bf k}_1,{\bf k}_2,{\bf q}} 
V_{{\bf k}_1,{\bf k}_2,{\bf q};\alpha\beta\gamma\delta}  
c^{\dagger}_{{\bf k}_1+{\bf q} \alpha} 
c^{\dagger}_{{\bf k}_2-{\bf q} \beta} 
c_{{\bf k}_2 \gamma}
c_{{\bf k}_1 \delta} 
\nonumber\\
&=& \frac{1}{2\Omega} \sum_{{\bf q}}  
\left[g_0({\bf q}) \rho^{\mu}_{{\bf q}}
\rho^{\mu}_{- {\bf q}} + v_{\rm F}^{-2} g_{1ij}({\bf q}) 
J^{\mu}_{i {\bf q}} J^{\mu}_{j -{\bf q}} \right],  
\label{hint}
\end{eqnarray}
where the spin density and spin current operators are defined as
\begin{eqnarray}
\rho^{\mu}_{\bf q} =  
\frac{\hbar}{2} \sum_{{\bf k}} c^{\dagger}_{{\bf k} \alpha}
\sigma^{\mu}_{\alpha \beta}   
c_{{\bf k}+{\bf q} \beta}, 
\nonumber
\end{eqnarray}
and 
\begin{eqnarray}
{\bf J}^{\mu}_{\bf q}&=& \frac{\hbar^2}{2}
\sum_{{\bf k}} \frac{2 {\bf k} + {\bf q}}{ 2 m_e}
c^{\dagger}_{{\bf k} \alpha}  
\sigma^{\mu}_{\alpha \beta}  c_{{\bf k}+{\bf q} \beta},  
\nonumber
\end{eqnarray}
respectively ($ \Omega $ denotes the volume; we sum over repeated
indices.). 
The first term in Eq.(\ref{hint}) describes 
the paramagnon exchange and $g_0({\bf q})$ is given by   
\begin{equation}
g_0({\bf q}) =-\frac{I({\bf q})}{1 - I({\bf q}) \chi ({\bf q})}
\end{equation}
where $I({\bf q})$ is the spin exchange interaction ($ I({\bf q}) =
I_0 \{ 1 + c (q/2k_F)^2 \} $ with $ I_0 $ correspond to the spin-spin
contact interaction). 
The spin susceptibility $\chi ({\bf q})$ is defined as 
$\langle T_{\tau} \rho^{\mu}_{\bf q}\rho^{\nu}_{-{\bf q}}\rangle_{\rm 1PI}  
=\chi ({\bf q}) \delta^{\mu \nu}$ where 
$\langle T_{\tau} \rho^{\mu}_{\bf q}\rho^{\nu}_{-\bf q}\rangle_{\rm
  1PI}$ denotes the one-particle irreducible (1PI) diagrams 
for the static spin-spin correlation in the normal state, 
and $\left<T_{\tau} \cdot\cdot\cdot \right>$ is the thermal
expectation value of operators with the imaginary time-ordered
product. The single-loop diagram in our two-dimensional model leads to 
a constant susceptibility $ \chi({\bf q}) = \chi_0 $ for $ q
< 2 k_F $ so that the $q$-dependence of $ g_0 ({\bf q}) $ occurs through
$ I({\bf q}) $ for small $ q $. 

The second term in Eq. (\ref{hint}) results from the 
spin current-spin current contact interaction, $ -(I_1/2 v^2_F)
{\bf J}^{\mu} ({\bf x}) \cdot {\bf J}^{\mu} ({\bf x}) $.  
Including the polarization analogous to the above case, we obtain
$ g_{1ij}({\bf q})=g_1({\bf q})(\delta_{ij} - q_i q_j / q^2),  
$ where 
\begin{equation}
g_1({\bf q})= - \frac{I_1}{1 - (I_1/ v_{\rm F}^2) \chi_c({\bf q})}. 
\end{equation}
$\chi_c({\bf q})$ is defined via the relation with 
the 1PI static spin current-spin current correlation as 
$\left< T_{\tau} J^{\mu}_{i{\bf q}} 
J^{\nu}_{j-{\bf q}}\right>_{\rm 1PI}=\chi_c ({\bf q}) 
\delta^{\mu \nu}(\delta_{ij} - q_i q_j / q^2)$. For small $q$  we
find\cite{comment0} , 
$$
\chi_c ({\bf q}) \simeq v_{\rm F}^2 (1 + \frac{q^2}{4 k_{\rm F}^2}) 
\chi ({\bf q}) .
$$

The structure of the interaction and the coupling constants are
derived from a  
short-ranged repulsive two-particle interaction (e.g. Coulomb for
charged particles),
\begin{equation}
{\cal H}_{c}= 
\frac{1}{2 \Omega} \sum_{{\bf k},{\bf k}',{\bf q}} \sum_{\alpha, \beta} 
U ({\bf q}) c^{\dag}_{{\bf k} + {\bf q}, \alpha} c_{{\bf k}, \alpha}  
c^{\dag}_{{\bf k}'-{\bf q}, \beta} c_{{\bf k}', \beta}.
\end{equation}
Using the SU(2)-identity $ 3 \delta_{\alpha \delta} \delta_{\beta
  \gamma} 
= 2 \sigma^{\mu}_{\alpha \beta} \sigma^{\mu}_{\gamma \delta} +
  \sigma^{\mu}_{\alpha \delta} \sigma^{\mu}_{\gamma \beta} $ 
and neglecting the irrelevant terms (which show interactions for the
  spin singlet pairing channel) we obtain
  a spin-spin exchange interaction as well as spin current-spin
  current interaction with the above coupling constants,
\begin{eqnarray}
I_0 &=& \frac{4}{3 \hbar^2} U({\bf q})|_{{\bf q}=0}, 
\quad c = \frac{2 I_1}{I_0} \sim  \frac{(2 k_{\rm F} l)^2}{3}, 
\nonumber\\
I_1 &=& -  \frac{1}{3} \frac{2}{3 \hbar^2} (2 k_{F})^2 \frac{\partial^2 U ({\bf q})}{\partial q_i^2}|_{{\bf q}=0} 
\sim \frac{2}{3} (k_{\rm F} l)^2 I_0 
\end{eqnarray}
where $ l $ is range of the interaction $ V_c $ (Thomas-Fermi
screening length for Coulomb interaction). 

Using Eq.(\ref{hint}) and the  BCS decoupling scheme we derive the  
following self-consistent equations in the weak-coupling limit for
the $p$-wave pairing channel, which we assume to be dominant 
\begin{eqnarray}
d_{\mu}(\bf{k})&=& - \frac{1}{\Omega}
\sum_{{\bf k}^{\prime}}\frac{\hbar^2}{4} 
\left[
\tilde{g}_0 - \tilde{g}_1 
 \right] \hat{\bf k} \cdot \hat{\bf k}^{\prime} 
\frac{d_{\mu}({\bf k}^{\prime})}{2 E_{\bf k}^{\prime}}
\tanh \frac{E_{\bf k}^{\prime}}{2 k_B T}, 
\label{gap-eq}
\end{eqnarray}
where the wave vectors $ {\bf k} $ and $ {\bf k}' $ are restricted
to certain range close to  
the Fermi surface ($ |{\bf k}|=|{\bf k}'|= k_F $), defined by a
cutoff energy $ \epsilon_c >
|\epsilon_{\bf k}|,|\epsilon_{\bf k}'| $ ($ d_{\mu}({\bf k})=
d_{\mu j} \hat{k}_j $). The quasiparticle
energy is $E_{\bf k}=\sqrt{\epsilon_{\bf k}^2 
+ (1/2) tr[\Delta_{\bf k}^{\dagger} \Delta_{\bf k}]}$. The
effective coupling
constants for the $p$-wave channel 
are obtained from Eq.(\ref{hint}) through
\begin{equation}
\tilde{g}_0 =  \langle g_0({\bf k} -{\bf k}') \hat{{\bf k}} \cdot
\hat{{\bf k}}'  
\rangle \quad \mbox{and} \quad \tilde{g}_1 = (1/2)\langle  g_1({\bf k} -{\bf k}') \rangle 
\end{equation}
where the average is taken over the Fermi surface for both wave
vectors ($ |{\bf k}|=|{\bf k}'|=k_F$). The constant $\left[
\tilde{g}_0 - \tilde{g}_1 \right]$ is negative
so that we find a solution for $p$-wave superconductivity in
Eq.(\ref{gap-eq}). It is easy to see that both types of pairing states 
(A- and B-phase) satisfy the identical self-consistent equation. 

Let us now consider the correction to the coupling functions $g_0({\bf q})$
 and $g_1({\bf q})$ below T$_c$, 
\begin{eqnarray}
g_0({\bf q}) \delta^{\mu \nu} &\rightarrow& 
g_0({\bf q}) \delta^{\mu \nu} + \delta g_0^{\mu \nu}({\bf q}), 
\nonumber\\
g_{1ij}({\bf q}) \delta^{\mu \nu} &\rightarrow& 
g_{1ij}({\bf q}) \delta^{\mu \nu} + \delta g_{1ij}^{\mu \nu}({\bf q}).  
\end{eqnarray}
It is sufficient to restrict ourselves to temperatures close to $ T_c
$ so that we can restrict to the lowest order contributions of the
order parameters \cite{FOOT}. Thus, we find,
\begin{eqnarray}
\delta g_0^{\mu \nu}({\bf q})&=&\left\{g_{0}({\bf q})\right\}^2 \delta 
\pi^{\mu \nu}_{00}({\bf q})
\nonumber\\
&=&C \left\{g_{0}({\bf q})\right\}^2 f({\bf q})  
\left\{ \delta_{\mu \nu} d^*_{\rho l} d_{\rho l}  
- 2 {\rm Re} d^*_{\mu l} d_{\nu l}\right\}
\\
\delta g_{1ij}^{\mu \nu} ({\bf q})&=&\left\{g_{1}({\bf q})\right\}^2 
 \delta \pi_{ij}^{\mu \nu} ({\bf q}) / v_{\rm F}^2 
\nonumber\\
&=&C \left\{g_{1}({\bf q})\right\}^2(\delta_{ij} - \frac{q_i q_j}{q^2}) 
h({\bf q})  \nonumber\\
&&\times\left\{ \delta_{\mu \nu} d^*_{\rho l} d_{\rho l} 
- 2 {\rm Re} d^*_{\mu l} d_{\nu l} 
\right\}.
\end{eqnarray}
where $\delta \pi^{\mu \nu}({\bf q})$ and 
$\delta \pi^{\mu \nu}_{ij}({\bf q})$ are the correction to  
the spin-spin correlation function and spin
 current-spin current correlation function, respectively.  
Here $C= \chi_0 / (4 \pi k_{B}^2 T_{c}^2) $, and 
\begin{eqnarray}
f({\bf q})&=&1 / \sqrt{1 + \xi^2 q^2}, 
\nonumber\\
h({\bf q})&=&(1 + q^2 / 4 k_{\rm F}^2) f({\bf q}).   
\nonumber
\end{eqnarray}
with $\xi = v_{\rm F} / 2 \pi k_{\rm B} T_{\rm c}$. Note that the
approximative analytic form of the function $f({\bf q}) $ can be
obtained in an analogous way as discussed in Ref.\cite{Goryo-Sigrist}.   
The corrections are calculated by using the anomalous Green function
linearized in the gap function: 
${\cal F}_{\alpha\beta}({\bf k}, i \omega_m)=i \Delta({\bf k})_{\alpha\beta} 
/ (\omega_m^2 + \epsilon({\bf k})^2) $.  
The modification of the coupling constant in Eq. (\ref{gap-eq}) is 
\begin{equation}
\delta (\tilde g_0^{\mu \nu} + \tilde g_1^{\mu \nu}) 
=C \gamma^{\rm sf} \left\{ \delta_{\mu \nu} d^*_{\rho l} d_{\rho l} - 
2 {\rm Re} d^*_{\mu l} d_{\nu l}\right\}
\label{sf-fb}
\end{equation}
with 
\begin{eqnarray}
\gamma^{\rm sf}&=&\langle  f({\bf k} - {\bf k}^{\prime}) 
\left\{g_{0}({\bf k} - {\bf k}^{\prime} )\right\}^2 
\hat{{\bf k}} \cdot \hat{{\bf k}}^{\prime} 
\nonumber\\
&&+ \frac{1}{2} h ({\bf k} - {\bf k}') 
\left\{g_{1}({\bf k} - {\bf k}')\right\}^2  
\rangle_{|{\bf k}|=|{\bf k}'|=k_F} \\
& \sim & \frac{\sqrt{2}}{k_{\rm F} \xi} \ln(k_{\rm F} \xi) \left[
    (g_0(0))^2 + \frac{1}{2}(g_1(0))^2 \right],
\nonumber
\end{eqnarray}
where we have used that the function $ f(q) $ is small for $ q \gg
\xi^{-1} \gg k_{\rm F} $. 
The effect depends on the order parameter structure via 
$\left\{ \delta_{\mu \nu} d_{\rho l}^* d_{\rho l} 
- 2 {\rm Re} d^*_{\mu l} d_{\nu l}\right\}$ which is selective for
different pairing states.

Now we turn to a different process contributing to the feedback effect
which is based on
the anomalous coupling of spin density-spin current leading an
interaction of the form 
$
H_{\rm an}= \Omega^{-1} \sum_{{\bf q}} \delta g_{0i}^{\mu \nu} ({\bf q}) 
\rho^{\mu}_{{\bf q}} J^{\nu}_{i - {\bf q}} / v_{\rm F}.
$
Such a coupling cannot exist in the normal state but only occurs as
a result of broken time reversal symmetry and parity, as is the case
for the chiral A-phase in Eq.(\ref{phases}). 
The coupling function $\delta g_{0i}^{\mu \nu} ({\bf q})$ to lowest order 
of the order parameter has the form 
\begin{eqnarray}
\delta g_{0i}^{\mu \nu} ({\bf q})&=&\left\{g_0({\bf q}) g_1({\bf q})\right\} 
\delta \pi_{0i}^{\mu \nu}({\bf q}) / v_{\rm F}
\nonumber\\  
&=& C \left\{g_0({\bf q}) g_1({\bf q})\right\} 
i \epsilon_{ij} q_i  f({\bf q}) / k_{\rm F}
\nonumber\\
&&\times\epsilon_{kl}\left\{\delta_{\mu \nu} {\rm Im} d^*_{\rho k} d_{\rho l}
 -  2{\rm Im} d^*_{\mu k} d_{\nu l}\right\} 
\label{sd-sc}
\end{eqnarray}
where $\delta \pi_{0i}^{\mu \nu}({\bf q})= 
\langle T_{\tau} \rho^{\mu}_{{\bf q}} J^{\nu}_{i - {\bf q}}\rangle$ 
shows the spin-spin current correlation function in the
superconducting state, calculated by using the linearized anomalous
Green function. Analogous to Ref.\cite{Goryo-Sigrist}, 
we find that the effective interaction for $p$-wave pairing channel 
has the form,
\begin{eqnarray}
V^{\rm an}_{{\bf k},{\bf k}^{\prime}, \alpha\beta\gamma\delta}&=&
\frac{\hbar^2 C}{4} 
(\sigma^{\mu})_{\alpha\delta}(\sigma^{\nu})_{\beta\gamma} 
\gamma^{\rm an} \left\{i \hat{\bf k}\times\hat{\bf k}'\right\}
\nonumber\\
&&\times \epsilon_{kl}\left\{ \delta_{\mu \nu} 
{\rm Im} d^*_{\rho k} d_{\rho l}
-  2{\rm Im} d^*_{\mu k} d_{\nu l}\right\}, 
\end{eqnarray}
where 
\begin{eqnarray}
\gamma^{\rm an}&=&
\left<g_{0}({\bf k} - {\bf k}^{\prime}) 
g_{1}({\bf k} - {\bf k}^{\prime}) f({\bf k} - {\bf k}^{\prime}) 
\right>_{|{\bf k}|=|{\bf k}|^{\prime}=k_{\rm F}} 
\nonumber\\
&\sim & \frac{\sqrt{2}}{k_{\rm F} \xi} \ln(k_{\rm F} \xi) g_0(0)
  g_1(0). 
\end{eqnarray}
Note that the dependence on the order parameter is different from that
in Eq.(\ref{sf-fb}) and that indeed only the chiral pairing state 
generates this contribution. 

The total feedback contribution to be added to the $p$-wave gap
equation (\ref{gap-eq}) is given by
\begin{eqnarray}
\delta d_{\mu i}&=& 
-\frac{1}{\Omega}\sum_{{\bf k}^{\prime}}\frac{\hbar^2 C}{4}
\left[ \gamma^{\rm sf} \delta_{ij}
\left\{ \delta_{\mu \nu} d_{\rho l}^* d_{\rho l} 
- 2 {\rm Re} d^*_{\mu l} d_{\nu l}\right\} 
\right. 
\nonumber\\
&&\left. - i \gamma^{\rm an} \epsilon_{ij} 
\left\{\delta_{\mu \nu} {\rm Im} (\epsilon_{kl} d^*_{\rho k} d_{\rho l})  
- 2{\rm Im} (\epsilon_{kl} d^*_{\mu k} d_{\nu l}) 
  \right\} \right]
\nonumber\\
&&\times\frac{d_{\nu j}}{2 E_{{\bf k}^{\prime}}}
\tanh \frac{E_{{\bf k}^{\prime}}}{2 k_B T}, 
\label{fb-gap}
\end{eqnarray}
Note that both  
$\gamma^{\rm sf}$ and $\gamma^{\rm an}$ are positive and  
are strong coupling corrections as indicated by 
the factor $ (k_F \xi)^{-1} $ ($ \ll 1 $).  

The relative magnitude of the two feedback contributions depends on
the parameter $ k_F l $, i.e. on the range of the interaction $ U
$. 

\begin{equation}
\frac{\gamma^{an}}{\gamma^{sf}} \sim \frac{g_0(0) g_1(0)}{g_0(0)^2 +
  g_1(0)^2 / 2} \sim (2 k_{\rm F} l)^2 (1 - I_0 \chi_0)
\label{sqhe-contribution}
\end{equation}
where the last analytic form is valid for $ k_{\rm F} l $ much smaller than
1. Since, however, the range of the interaction can be comparable with
the average interparticle distance, this ratio could of order 1 as
well and the anomalous contribution can be comparable to the ordinary
spin fluctuation feedback. In Fig. 1, we plot $I_0 \chi_0$ dependence of
the ratio $\gamma^{an}/\gamma^{sf}$ in the case $k_{\rm F} l \sim 1$ .

The feedback effect enters as a correction to the fourth-order terms
in the Ginzburg-Landau free energy. These terms are readily obtained from
Eq.(\ref{fb-gap}) as     
\begin{eqnarray}
\Delta F^{\rm fb}&=& \Gamma^{\rm sf}{\rm Re}( d_{\mu i}^* d_{\nu i}) 
\left\{ \delta_{\mu \nu} d_{\rho l}^* d_{\rho l} 
- 2 {\rm Re} (d^*_{\mu l} d_{\nu l}) \right\}
\nonumber\\
&&+ \Gamma^{\rm an}{\rm Im} (\epsilon_{ij} d^*_{\mu i} d_{\nu j} )
\left\{\delta_{\mu \nu} {\rm Im} (\epsilon_{kl} d^*_{\rho k} d_{\rho l})\right.
\nonumber\\
&&\left.- 2{\rm Im}( \epsilon_{kl} d^*_{\mu k} d_{\nu l})
  \right\}.  
\label{free-en}
\end{eqnarray}
with 

\begin{eqnarray}
\Gamma^{\rm sf} &=& \frac{C}{4 \hbar^2}\frac{\gamma^{sf}}{(\tilde{g}_0 - \tilde{g}_1)^2}  \\
\Gamma^{\rm an} &=& \frac{C}{4 \hbar^2}\frac{\gamma^{an}}{(\tilde{g}_0 - \tilde{g}_1)^2} 
\end{eqnarray}

As we mentioned previously, the part of $\Gamma^{\rm sf}$ is
equivalent to the standard spin fluctuation feedback as discussed in
literature \cite{HELIUM,KORSH,ABS-Kuroda}. As is
well known it modifies the condensation energy to stabilize the
(chiral) A-phase in Eq.(\ref{phases}) analogous to superfluid $^3$He. 
The anomalous contribution due to spin density-spin current coupling
in the $\Gamma^{\rm an} $-term yields an additional bias towards the
same A-phase. Considering the order parameter dependence in
Eq.(\ref{free-en}), we see that the feedback benefit only
occurs in the chiral condensate.  
This chiral correction is intimately related to SQHE in the chiral
superconductors or superfluid 
\cite{YAKO,Fisher}. SQHE is an effect where a the spin current 
is induced transverse to the gradient of a Zeeman field, i.e. 
$ {\bf J}^{\mu} \cdot {\bf \nabla} B^{\mu} = 0 $ where $ \mu $ denotes the
spin component. This effect has been discussed for states like the
A-phase. 
It was shown that a Chern-Simons term exists in the effective action 
for the SU(2) gauge field $(A_0^{(\mu)}, {\bf A}^{(\mu)})$
\cite{Read-Green} (the time component $A_0^{(\mu)}$ corresponds to
the Zeeman field).   
The SU(2) Chern-Simons term has a bilinear part of the 
gauge field with one space-time derivative and totally 
anti-symmetric with respect to space-time indices, i.e. 
$\kappa \epsilon_{\alpha \beta \gamma}A_{\alpha}^{(\sigma)}\partial_{\beta}
A_{\gamma}^{(\sigma)}$\cite{Deser-Jackiw-Templeton}, where the
subscripts ($ \alpha,\beta,\gamma $) denote space-time indices. 
The coupling constant $\kappa$ corresponds to the transverse
conductance for the spin degree of freedom and is obtained form 
the spin-spin current correlation 
$\pi^{\mu \nu}_{0j}({\bf q})$, which plays 
an essential role to introduce the 
chiral correction (Eq.(\ref{sd-sc})), by calculating    
\begin{eqnarray}
\kappa&=& \left.
\frac{1}{2!2!} i \epsilon_{ij} 
\delta^{\mu \nu} 
\frac{\partial}{\partial q_i} \pi^{\mu \nu}_{0j}({\bf q})\right|_{{\bf q}=0}. 
\end{eqnarray}
It was shown that at zero-temperature 
$\kappa$ is universal $1 / 2 \pi$ in the units $\mu_B$, the 
Bohr magneton, in the chiral $p$-wave states in contrast to
the related charge Hall effect which is non-universal
\cite{zero-field-Hall-effect,Fisher,Read-Green}. 

In summary, we have analyzed the anomalous contribution to the spin
feedback effect in spin triplet superconductors or superfluid in a
two-dimensional system. This effect together with the standard spin
fluctuation feedback effect favors the chiral $p$-wave phase,
corresponding to the A-phase in Eq.(\ref{phases}), and lifts in this
way the weak-coupling degeneracy among several spin-triplet 
pairing states. Chirality plays a similar role in the charge
chiral feedback effect which is, however, much weaker than the spin
chiral feedback effect, because in the former the physical U(1) gauge
field mediates the modified interaction \cite{Goryo-Sigrist}. 
The degeneracy between the A- and B-phase like pairing states is
also lifted by spin-orbit coupling \cite{HELIUM,KORSH,NG}. Since this
yields a correction to the second order term in the free energy it is 
generally more decisive. Nevertheless, our discussion shows that 
in 2D systems where the feedback effect is the decisive mechanism to
lift the degeneracy the anomalous coupling between spin-density and
spin-current can give an sizeable contribution.

\acknowledgments

The authors are grateful to A. Furusaki, M. Kohmoto,
H. Kohno, T. Morinari, N. Nagaosa and M. Sato for helpful discussions.  
This work has been financially supported by a Grant-in-Aid of the
Japanese Ministry of Education, Science, Culture and Sports.

\newpage

\begin{figure}
\onefigure{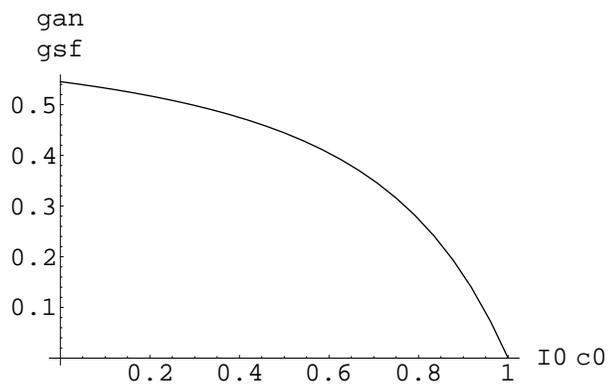}
\caption{$I_0 \chi_0$ dependence of the ratio $\gamma_{an} / \gamma_{sf}$
(Eq. (\ref{sqhe-contribution})) in the case $k_{\rm F} l \sim 1$. }
\label{f.1}
\end{figure}


\begin{thebibliography}{0}
\bibitem{FOOT} We also restrict our discussion to the static part
for simplicity. The inclusion of dynamical contributions does not
change the result qualitatively, but makes the analysis considerably
more complicated.
\end{thebibliography}
\end{document}